# Atom-by-atom Imaging of Moiré Phasons using Electron Ptychography

Yichao Zhang[1,4,5], Ballal Ahammed[2,4], Sang Hyun Bae[1,4], Chia-Hao Lee[1,4], Jeffrey Huang[1,4], Mohammad Abir Hossain[2,4], Tawfiqur Rakib[2,3], Arend van der Zande[1,2,3,4], Elif Ertekin[2,3,4], and Pinshane Y. Huang[1,3,4]*

[1]Department of Materials Science and Engineering, University of Illinois Urbana-Champaign; Urbana, Illinois 61801, USA.

[2]Department of Mechanical Science and Engineering, University of Illinois Urbana-Champaign; Urbana, Illinois 61801, USA.

[3]Materials Research Laboratory, University of Illinois Urbana-Champaign; Urbana, Illinois 61801, USA.

[4]The Grainger College of Engineering, University of Illinois Urbana-Champaign; Urbana, Illinois 61801, USA.

[5]Department of Materials Science and Engineering, University of Maryland; College Park, Maryland 20742, USA.

*Corresponding author. Email: pyhuang@illinois.edu

**Abstract:** Twisted 2D materials exhibit unique vibrational modes called moiré phonons, which arise from the moiré superlattice. Here, we demonstrate atom-by-atom imaging of phasons, an ultrasoft class of moiré phonons in twisted bilayer $WSe_2$. Using ultrahigh-resolution (<15 pm) electron ptychography, we image the size and shape of each atom to extract time-averaged vibrational amplitudes as a function of twist angle and position. We observe several signature properties of moiré phasons, such as increased vibrational amplitudes at solitons and AA-stacked regions. By correlating experiments with molecular dynamics simulations and lattice dynamics calculations, we show phasons dominate the thermal vibrations in low-angle twisted bilayers. These results represent a powerful route to image thermal vibrations at atomic resolution, unlocking experimental studies of a thus-far hidden branch of moiré phonon physics.

**Introduction**

Twisted van der Waals bilayers exhibit unique vibrational modes, broadly defined as moiré phonons (*1–5*), arising from the moiré superlattice. These modes tune with interlayer twist angle and couple strongly to important physical properties of twisted van der Waals materials ranging from charge transport to unconventional superconductivity (*3–19*). So far, experimental studies of moiré phonons have focused on interlayer breathing and shear modes in the range of tens of cm$^{-1}$, which can be accessed with low-wavenumber Raman spectroscopy (*20, 21*). However, 2D moiré materials are predicted (*3–5*) to host a second class of moiré phonons, referred to as phasons or ultrasoft shear modes, which have never been directly observed.

At small twist angles of a few degrees, van der Waals bilayers reconstruct into moiré superlattices (*22–24*) containing quasi-periodic regions of rotationally aligned regions (such as AA or AB), separated by a network of stacking faults often referred to as solitons (Fig. 1A–C). In these low-twist-angle moirés, the solitons form a soft elastic network which hosts a unique set of vibrational modes that is not present in individual monolayers. The moiré phason modes are the in-plane vibrational modes of the soliton network: they correspond to localized interlayer sliding at the soliton domain walls and translations of the moiré superlattice (*3–5*). Such modes have been referred to in the literature as ultrasoft shear modes, ultrasoft moiré phonons, or phasons. These terms are often used interchangeably, even though phasons are more precisely defined as vibrational modes of incommensurate phases, resembling acoustic modes with zero-energy excitations at infinitely long wavelengths (*25, 26*). We follow the convention of referring to the ultrasoft shear modes as phasons regardless of their precise twist angle and commensurability (*27*), because real moiré superlattices are typically disordered across length scales of tens of nm and above (*28–32*).

While moiré phasons have been discussed in several theoretical and computational works (*3–5*, *9*, *25*, *27*, *33–35*), they have only been detected indirectly through electron-phason coupling (*36*). These phasons are extremely difficult to access experimentally because their predicted frequencies are very low, on the order of 1 cm$^{-1}$ or 0.1 meV (*5*), placing them out of reach of conventional spectroscopic techniques. Yet moiré phasons have already been implicated in diverse and important phenomena: they have been predicted to underlie superlubricity and disorder in the moiré lattice (*27*), alter charge transport (*33*, *37*) and exciton diffusion (*9*), dominate thermal conductivity (*37*) and specific heat at low temperatures (*5*), and couple strongly to the emergent electronic properties at low angle twists (*4*, *27*, *36*, *38*).

The challenges in measuring moiré phasons point to a gap in experimental techniques for probing thermal vibrations. Existing diffraction-based measurements, such as X-ray or neutron scattering, lack the sensitivity and atomic-scale spatial resolution needed to directly visualize phasons. On the other hand, spectroscopic techniques can either access sufficiently low frequencies to detect moiré phonon modes (e.g., Raman spectroscopy) (*1*, *21*) or operate at atomic-scale resolution (*e.g.* monochromated energy-loss electron spectroscopy) (*39–42*), but not both.

Rapid advances in the spatial resolution of electron ptychography offer a potential new route for measuring atomic vibrations. Electron ptychography can achieve spatial resolutions of tens of picometers (*43–52*), sufficiently high that thermal vibrations measurably blur atomic images (*44*). These advances indicate the potential for ptychography to serve as a real-space, atomically resolved method for imaging thermal vibrations, with the potential to newly access low-frequency, anisotropic, and spatially localized modes that are beyond the reach of current methods.

## Achieving <15 pm resolution with multislice electron ptychography

Figure 1 shows how we use multislice electron ptychography (MEP) to achieve the picometer-scale spatial resolution required to image thermal vibrations in twisted bilayer $WSe_2$. We fabricate a $WSe_2$ bilayer (see Materials and Methods) with a 1.7° interlayer twist angle (0° is defined as parallel or AB stacking). Next, we image the sample with aberration-corrected annular dark-field scanning transmission electron microscopy (ADF-STEM) in order to visualize the moiré superlattice and locate a region of interest. Fig. 1A–C show the resulting images and corresponding atomic structures, which reveal the reconstructed moiré superlattice and its network of soliton stacking faults.

Fig. 1D shows an image of the bilayer $WSe_2$ obtained using MEP. MEP (*44*, *53*) calculates an object from a series of diffraction patterns acquired as an atomic-scale electron beam is rastered across a sample. In an ideal case, the reconstructed phase object is proportional to the electrostatic potential, which is blurred by the thermal vibration of atoms during the few-second image acquisition. MEP has received intense recent interest because it has achieved the highest spatial resolution of any real-space imaging method (*44*, *54*). With MEP (Fig. 1D), we obtain exceptionally clear images of each atomic site. For example, the Fourier transform of the ptychography image in Fig. 1D displays peaks indicating 0.32 Å information transfer, triple that of the aberration-corrected STEM image in Fig. 1C (0.96 Å). We further improve the 2D spatial resolution of MEP by using an extended depth of field (EDF) algorithm (*55*, *56*) to extract the 2D projection in Fig. 1E (see Materials and Methods). EDF algorithms are commonly used to extract all-in-focus image projections from focal series (*56*); here we adapt this method to remove background signals from MEP depth slices that do not contain atoms. We apply this method to all ptychography images unless otherwise specified. Using this method, we improve the information transfer even further to < 0.29 Å. This high resolution allows us to resolve detailed atomic structures, even in atom pairs that are blurred together in conventional MEP projections (blue boxes in Fig. 1G–H). As a second measure of resolution, we measure the minimum projected atomic spacings resolvable with a visible intensity dip between the atoms, in analogy to the Abbe limit; we observe atomic spacings of 19.6 pm (Fig. 1F) and 14.7 pm (Fig. S1). These values approach calculated room temperature in-plane vibrational amplitudes of 5–6 pm for monolayer $WSe_2$ (*57*), indicating that the vibrational amplitudes might be accessible with ptychography.

**Visualization of moiré phasons in a soliton domain wall**

In Figure 2, we leverage the extremely high spatial resolution of ptychography to measure anisotropic thermal vibrations in a 2D moiré superlattice. We acquire and reconstruct an MEP image of 1.7°-twisted bilayer $WSe_2$ at room temperature, in a region containing a soliton and AB domains (Fig. 2A). As shown in Fig. 2B–C, we fit 2D Gaussians to each atomic site in the image, obtaining their in-plane short- and long-axis Gaussian root mean square (RMS) widths, ellipticity, and long-axis orientation (see Materials and Methods). These data are akin to atomic anisotropic displacement parameters (*58*), also called thermal ellipsoids, obtained from the Debye-Waller factor in X-ray diffraction. Unlike X-ray measurements, however, our method is atomically resolved rather than spatially averaged, making it ideal for resolving spatially localized modes such as moiré phonons and phasons.

Fig. 2D–E shows an atom-by-atom map and corresponding polar plot displaying the magnitude and orientation of elliptical 2D Gaussian fits of the single W atoms (see Supplementary Information). These data show that within the soliton, W atoms have increased long-axis lengths

and align parallel to the soliton direction (-36° in Fig. 2E), whereas in the AB region, their orientations are shorter with more randomly distributed orientations. A corresponding analysis for Se columns shows similar behavior (Fig. S3). As a test of robustness, we repeat our processing methods on conventional MEP projection images without applying EDF and obtain similar results (Fig. S4). The large amplitude vibrations, their strong ellipticity, and their localization to the solitons are unusual features that are consistent with the predictions for moiré phasons (*3–5*).

We perform harmonic lattice dynamics calculations (see Methods) to visualize the phason modes and compare them to our experimental data. Fig. 2F–G and Supplementary Video 1 show the lowest energy eigenmode of a 1.7° moiré superlattice, represented as displacement vectors for the two W layers (see Fig. S5 for the eigenvector maps of all atoms). In this phason mode, the motion is highly localized to the unstable stacking regions: the solitons (orange) and their connection points, the AA sites (blue). Within the solitons, the W and Se atoms in each monolayer displace in parallel, whereas the two $WSe_2$ layers displace antiparallel with respect to one another (Fig. 2H). The second lowest energy phason mode, nearly degenerate at the gamma point, contains similar motions for the symmetry-equivalent solitons in the moiré superlattice (Fig. S5 and Supplementary Video 2). The dispersion relations of these modes are shown in Fig. S6 and discussed in supplementary text. Our experimental ellipticity maps in Fig. 2D are qualitatively consistent with the anisotropic thermal vibration expected to arise if the phason modes dominate the vibrational amplitudes. These data indicate that our ptychography methods can directly image the moiré phason modes and their impact on thermal vibrations.

**Thermal vibrations across the moiré superlattice**

In Figure 3, we investigate how phasons and higher-frequency moiré phonons impact the vibrational motions and atomic profiles in MEP, considering all thermally activated modes. We study a 2.45°-twisted bilayer WSe$_2$, as shown in the MEP image in Fig. 3A. Our experimental measurements are conducted at room temperature, and so while we expect the vibrational amplitudes to be dominated by lowest frequency and thus highest amplitude modes, multiple phonon modes should be populated according to a Boltzmann distribution. We conduct molecular dynamics (MD) simulations to understand the total atomic vibration amplitudes from all thermally activated phonon modes at room temperature (see Materials and Methods). First, we used MD to calculate the atomic trajectories at 300 K. The atomic coordinates at each time step ($\Delta t = 0.5$ fs for a total of 3 ns, corresponding to ~30 periods of the lowest frequency modes) populate a probability density function (PDF) for each atom. Fig. 3B shows these probability maps and corresponding 2D Gaussian fits for three W atoms in the AA, AB, and soliton regions. These PDFs reveal key trends in the real-space distribution of moiré phonon modes. The total vibrational amplitudes are isotropic in AA and AB regions (approximately equal short- and long-axis amplitudes with $\sigma_{AA}>\sigma_{AB}$), and anisotropic in the soliton, consistent with the behavior observed in Figure 2.

Next, we visualize the varying vibrational amplitudes in real space, comparing simulation and experiment. Fig. 3C,E display experimental and MD maps showing the short-axis amplitudes of the thermal ellipsoids for individual W atoms, comparing similar fields of view for a 2.45°-twisted superlattice. We focus first on the short axes of the W atoms because we expect them to be more robust to potential experimental artifacts, such as errors in 2D Gaussian fitting. To produce the experimental map in Fig. 3C, we acquire, reconstruct, and fit 2D Gaussians to each W atom in MEP data as described previously. Next, we isolate the thermal vibration amplitudes from the ptychographic atom profiles by deconvolving a blurring factor, obtained from simulations, to remove the contributions from the finite width of the atomic potential and the finite resolution of the ptychographic reconstruction. See Fig. S7 and Materials and Methods for details of the ptychography simulations and deconvolution, as well as extraction of thermal root mean square displacement (RMSD) from MD simulation results. We obtain remarkable agreement between the experimental data in Fig. 3C and the MD simulations in Fig. 3E; both show the highest amplitudes near the AA regions, followed by AB and soliton regions. The large vibrational amplitudes at AA sites can be understood intuitively: as discussed previously, the AA sites are nodes of the elastic network of solitons, where vibrations are expected to be the largest. In addition, the AA sites are energetically unfavorable—they are local maxima in the interlayer potential energy landscape, a property that should amplify atomic vibrations.

We further confirm that our experimental signals arise from phasons and higher-frequency moiré phonons by duplicating our analyses on simulated ptychography data with varying phonon symmetries. First, we conduct ptychography simulations on $WSe_2$ bilayers with isotropic frozen phonons, as used in standard STEM simulations (*59*). In the isotropic frozen phonon model, atoms are given random displacements according to a Gaussian probability distribution with a specified RMSD; because these simulations lack the detailed spatial and frequency behavior of real phonon modes, they are useful controls to validate phonon detection. We compared these isotropic ptychography simulations against a more sophisticated model containing phonons calculated using MD, which include moiré phonons and phasons, as well as conventional lattice phonons (see Materials and Methods). As shown in Fig. S8, ptychography simulations with MD phonons replicate our experiment, exhibiting increased short-axis Gaussian RMS width near AA sites, while ptychography simulations with isotropic phonons do not. This correspondence rules out ptychography and Gaussian fitting artifacts as the source of our signals, indicating that our measured spatial distribution of atom sizes stems directly from local variations in phonon vibration amplitude associated with the local stacking orders within the moiré superlattice.

Next, we examine ptychography's ability to quantify the thermal vibration amplitude $\sigma$ and its spatial variations. Fig. 3D shows histograms comparing the vibrational amplitudes in different regions of the moiré superlattice for single W atoms, extracted from an experimental ptychography reconstruction. Using these distributions, we calculated mean W short-axis amplitudes of $\sigma_{AB} = 6.0 \pm 0.2$ pm in the AB regions, $\sigma_{soliton} = 5.8 \pm 0.3$ pm, and $\sigma_{AA} = 6.8 \pm 0.2$ pm; these reported errors are the standard error of the mean, which we use to assess our ability to determine shifts in $\sigma$ values within a single image; in addition, we estimate systematic errors of $\pm 1$–$2$ pm in $\sigma$ for each image due to uncertainties in the deconvolution of non-thermal contributions. For comparison, an analysis based on MD coordinates in a similar field of view yields $\sigma_{AB} = 5.5$ pm, $\sigma_{soliton} = 5.5$ pm, and $\sigma_{AA} = 5.6$ pm. These values are also consistent with literature values of the room-temperature, in-plane RMSD of W atoms (5.5 pm) calculated using first principles (*57*). Note: we obtain these averages from the image shown in Fig. 3A, which contains regions that approach AA stacking but do not reach the AA centers. These data show that, although the experimental amplitude measurements for single atoms are noisy and exhibit large spread, their mean values are sufficiently precise to show key trends: here, an increase in $\sigma_{short}$ near AA-stacked regions.

**Evolution of moiré phasons with twist angle**

Finally, in Figure 4 we investigate the evolution of moiré phason modes with twist angle and superlattice structure. Fig. 4A–C shows ptychography images of 1.7°, 2.45°, and 6.0°-twisted bilayers. These twisted bilayer samples span the different major regimes of moiré superlattices and their phonon behaviors: strongly reconstructed (<2°), a transitional regime (2°–6°), and near-rigid structures (>6°) where minimal lattice reconstructions occur (*5*, *21*). In order to highlight the evolution of the phason modes, we plot the average thermal vibration amplitude, $\sigma_{ave}$, for each twist angle from experiment (Fig. 4D–F) and MD (Fig. 4G–I). We use $\sigma_{ave}$ for each atom, rather than $\sigma_{short}$, as plotted in Figure 3, in order to better capture the anisotropic phasons. To generate these plots, we fit 2D Gaussians to single W atoms in each image, isolate the thermal contributions (see Materials and Methods), and then calculate a geometric average of the long- and short-axis thermal vibration amplitudes.

We observe localized, enhanced vibrations at the solitons and their connection points in both experiment (Fig. 4D–E) and MD (Fig. 4F–G). These trends are visible directly in the maps of $\sigma_{ave}$. For example, for the 1.7° twist, Fig. 4D,G show an increase in $\sigma_{ave}$ in the soliton compared to the surrounding AB region. For the 2.45° bilayer (Fig. 4E,H), the most visible increases in $\sigma_{ave}$ are observed near the AA stacked regions. In the 6.0° bilayer, the regional changes in vibration amplitude become more difficult to distinguish visually. There are likely two sources of the lack of visual trends in the 6.0° data. First, our MD simulations predict that the regional differences in the vibration amplitudes are smaller for larger twist angles (Fig. 4G,H,I). Second, for larger twist angles, the moiré unit cells shrink and the distinct stacking regions become smaller and more tightly spaced, making regional differences harder to discern.

Overall, these data are a good match with predictions for twist angle dependent changes in moiré phasons, as well as a broader understanding of how lattice reconstructions impact the interlayer interactions and moiré phonons. In the strongly reconstructed regime (<2°), the lattice reconstructions dominate, producing solitons and their corresponding phasons (*3–5*). In the transitional regime, where lattice reconstructions decrease (*21, 35*), the ultrasoft modes are predicted to gradually lose their localization to the soliton boundaries (*5*). For large twist angles, as the lattice reconstruction effects vanish, the structure and phonon behavior eventually approach that of two rigid $WSe_2$ layers (*5, 21, 60*).

**Summary and conclusions**

To conclude, we develop an atom-by-atom method to image thermal vibrations using electron ptychography and use it to reveal moiré phasons in twisted bilayer $WSe_2$. By doing so, we validate the signature predictions of ultrasoft moiré phasons: the emergence of spatially localized, anisotropic vibrations at the solitons and AA sites in low-angle moirés, features that were thought to be out of reach of experiments. Our work shows clearly that phasons dominate the thermal vibrations of reconstructed moiré structures, adding to the growing body of evidence that these ultrasoft modes play a key role in the properties of low-angle 2D moiré materials. More broadly, our electron ptychography methods introduce a unique and powerful way to measure spatially non-uniform vibrational modes, a capability that should prove impactful for understanding the vibrational properties of defects and interfaces.

**Acknowledgments:** Electron microscopy work was carried out in the Materials Research Laboratory Central Research Facilities at the University of Illinois and the NSF-MRSEC shared facilities. This work used NCSA Delta GPU at National Center for Supercomputing Applications at the University of Illinois through allocation MAT240032 from the National Science Foundation supported Advanced Cyberinfrastructure Coordination Ecosystem: Services & Support (ACCESS) program. Molecular dynamics and harmonic lattice dynamics calculations were performed on PSC Bridges-2 at the Pittsburgh Supercomputing Center provided by the same Advanced Cyberinfrastructure Coordination Ecosystem: Services & Support (ACCESS) program. Additional computational resources were provided by the Illinois Campus Cluster operated by the Illinois Campus Cluster Program. We thank Linmin Wang for developing the 2D Gaussian fitting procedure and Prof. Rafael Fernandes and Indrajit Maity for discussions of the theory of phasons.

**Funding:** National Science Foundation MRSEC awards DMR-1720633 and DMR-2309037 (PYH, AMvdZ, EE); Air Force Office of Scientific Research PECASE award AF FA9550-20-1-0302 (PYH); Packard Foundation Fellowship (PYH)

**Author contributions:** Conceptualization: PYH; Methodology: YZ, CHL, BA, TR, SHB, MAH; Investigation: YZ, BA; Software: YZ, BA, JH, TR; Formal analysis: YZ, BA; Visualization: YZ, BA; Resources: SHB, MAH; Funding acquisition: PYH, EE, AMvdZ; Project administration: PYH; Supervision: PYH; Writing – original draft: YZ, PYH, BA; Writing – review & editing: YZ, BA, SHB, CHL, JH, MAH, TR, AMvdZ, EE, PYH

**Competing interests:** The authors declare no competing interests.

**Data and materials availability:** All data will be made available upon publication

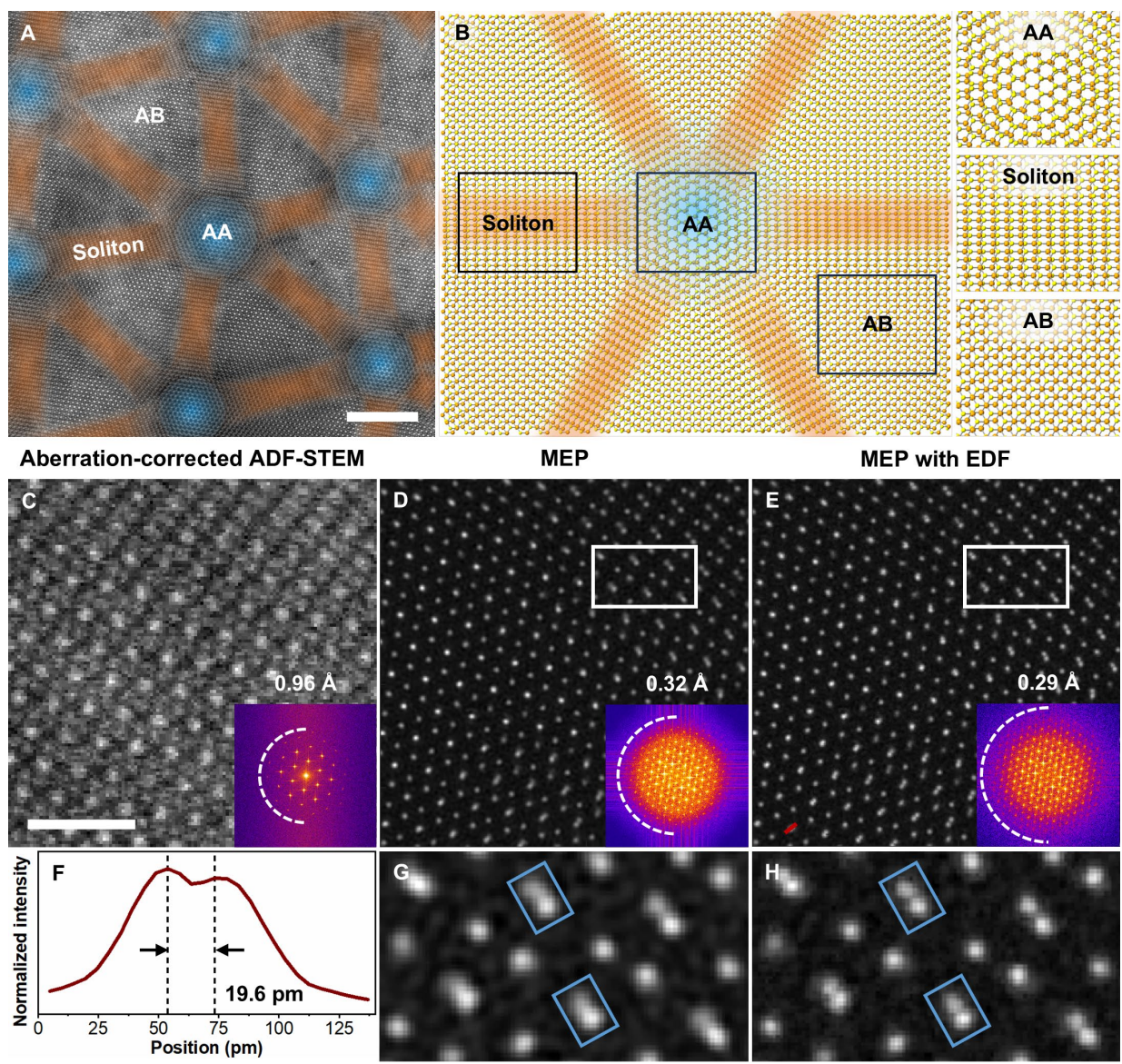

**Fig. 1 | Picometer-scale resolution achieved with multislice electron ptychography (MEP). A**, Aberration-corrected annular-dark-field scanning transmission electron microscopy (ADF-STEM) image of 1.7°-twisted bilayer $WSe_2$ with soliton (orange) and AA stacked (blue) regions shaded. Scale bar = 5 nm. **B**, Atomic structure of 1.7°-twisted bilayer $WSe_2$. Insets magnify representative AA, soliton, and AB regions. Orange and yellow colors label W and Se atoms, respectively. **C–E**, images of a similar region in 1.7°-twisted bilayer $WSe_2$ obtained by (**C**) Aberration-corrected ADF-STEM, (**D**) MEP, and (**E**) MEP with extended depth of field (EDF). Insets of **C–E** are their corresponding Fourier transforms. Information transfer limits (0.96 Å for aberration-corrected ADF-STEM, 0.32 Å for MEP and 0.29 Å for MEP with EDF) are labeled with dashed white arcs. Scale bar = 5 Å for **C–E**. Red line in **E** marks position of line profile in **F**. **F**, Line profile showing a resolvable projected atomic spacing of 19.6 pm. **G,H**, Magnified ptychographic phase images from areas outlined by white boxes in **D,E**, respectively. Blue rectangles label two pairs of atoms unresolvable with MEP but resolvable using MEP with EDF.

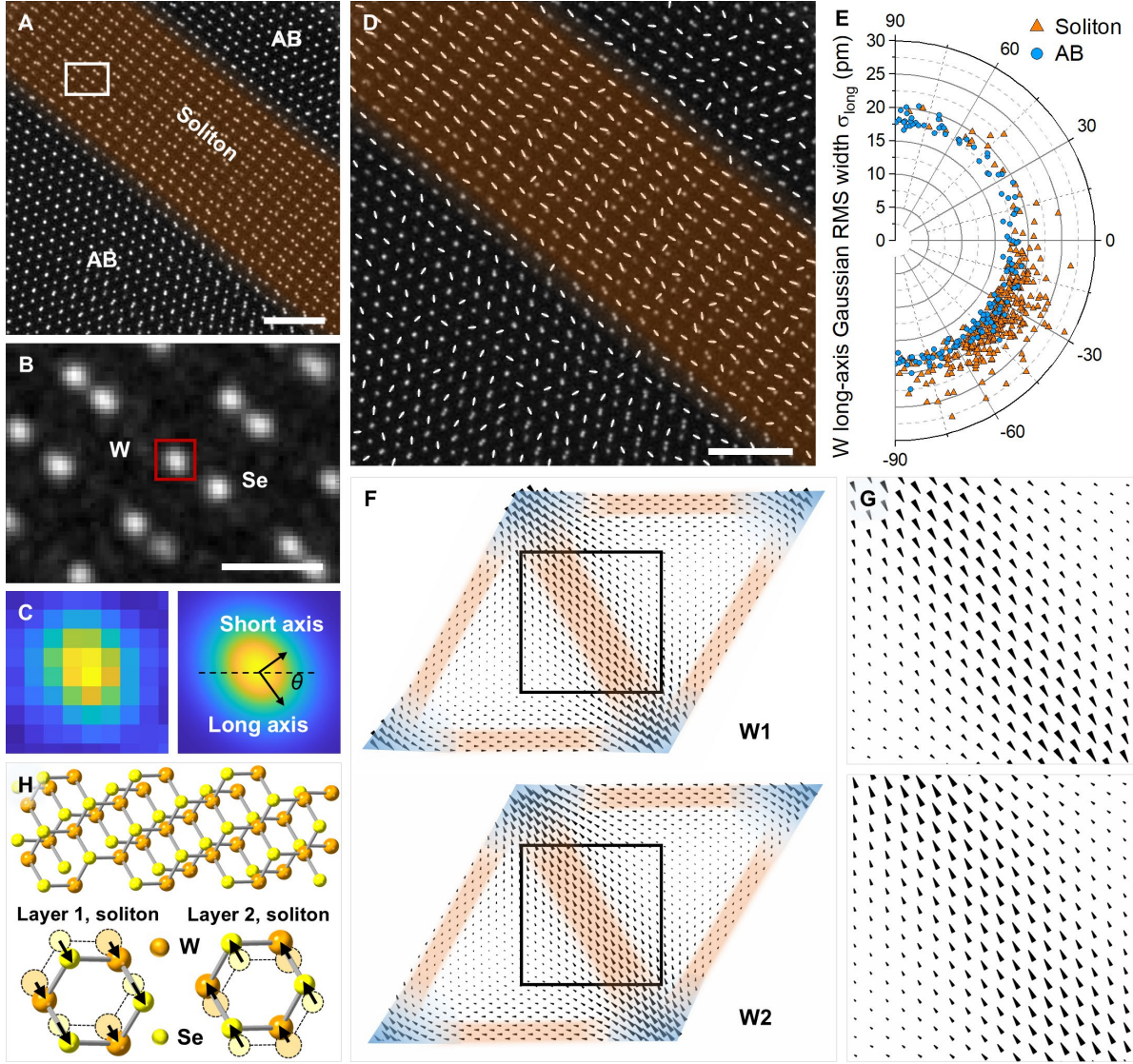

**Fig. 2 | Direct imaging of anisotropic phason modes at a soliton. A**, MEP image of 1.7°-twisted bilayer WSe$_2$. Orange shading marks the soliton region. **B**, Magnified image from white rectangle in **A** showing ellipticity of atoms. The ellipticity magnitude is calculated as $\sigma_{long}/\sigma_{short}$. Scale bar = 2 Å. **C**, Magnified image (left) and 2D Gaussian fit (right) of a single W atom labeled by the red box in **B**. **D**, Vector field map of anisotropic phonon vibration overlaid on **A**. White arrows indicate the relative magnitude and direction of the ellipticity of all single W atoms, magnified for visibility. Scale bar = 1 nm for **A**,**D**. **E**, Polar plot of single W atom long-axis Gaussian root-mean-square (RMS) width $\sigma_{long}$. **F**, Eigenvector maps of the lowest energy phonon mode of a 1.7°-twisted bilayer WSe$_2$ supercell. This mode is one of two phason or ultrasoft shear modes (see Fig. S5 for the second mode). Top and bottom panels respectively show the displacement of W atoms in the top and bottom layers. The phason mode is localized to the AA (blue) and soliton (orange) regions. Size of black arrows are magnified for visibility. **G**, Magnified eigenvector maps from the black boxes in **F**. **H**, Cartoon illustrations of atomic displacements in the soliton due to the phason mode. The top is the atomic structure of soliton. Bottom panels show the magnified in-plane atomic displacements in the different layers.

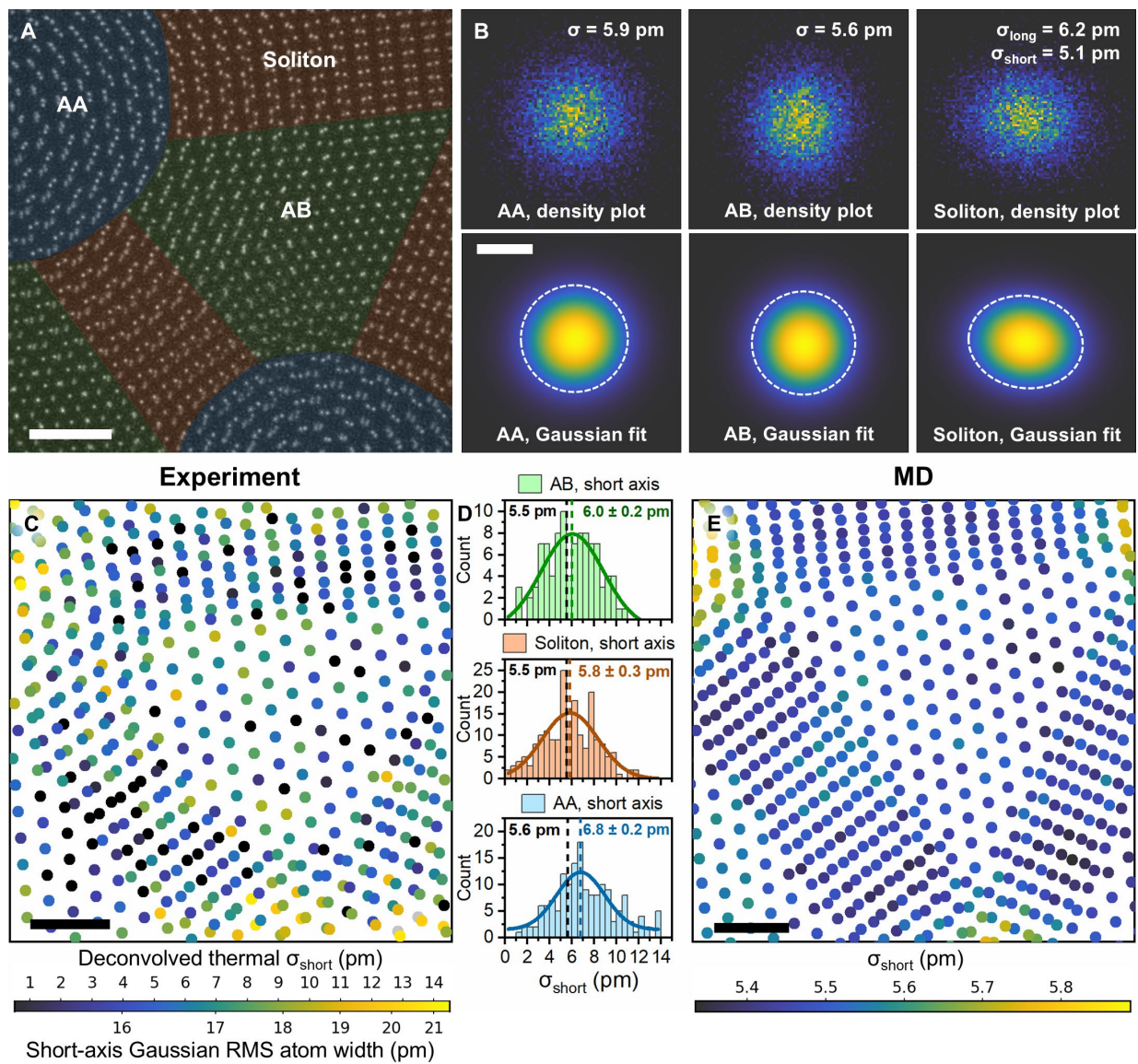

**Fig. 3 | Experimental and molecular dynamics (MD) comparison of single W atom sizes. A**, MEP image of 2.45°-twisted bilayer WSe₂. Soliton, AA, and AB regions are labeled with orange, blue, and green, respectively. **B**, Extraction of thermal vibration amplitude in 2.45°-twisted bilayer WSe₂ from MD simulations. Top row shows probability density function plots for single W atoms in AA (left), AB (middle), and soliton (right) regions. Bottom row panels show corresponding 2D Gaussian fits, where σ is the root mean square Gaussian width. See Materials and Methods for details on generating the PDFs and Gaussian fits. Dashed circles/ellipse mark full-width-at-quarter-maximum of fitted Gaussians. **C,E**, Experimental (**C**) and MD (**E**) short-axis thermal vibration amplitudes $\sigma_{short}$ of single W atoms. Bottom axis of color bar in **C** labels the corresponding short-axis Gaussian RMS atom width before deconvolution. **D**, Histograms of experimentally measured single W short-axis thermal vibration amplitude in regions AB (top), soliton (middle), and AA (bottom) regions, respectively. Bell curves label Gaussian fits. Vertical lines label centers of Gaussian fits for experiment (colored) and MD (black). Errors are standard errors of the mean. Scale bar = 1 nm for panels **A,C,E**. Scale bar = 10 pm for all panels in **B**.

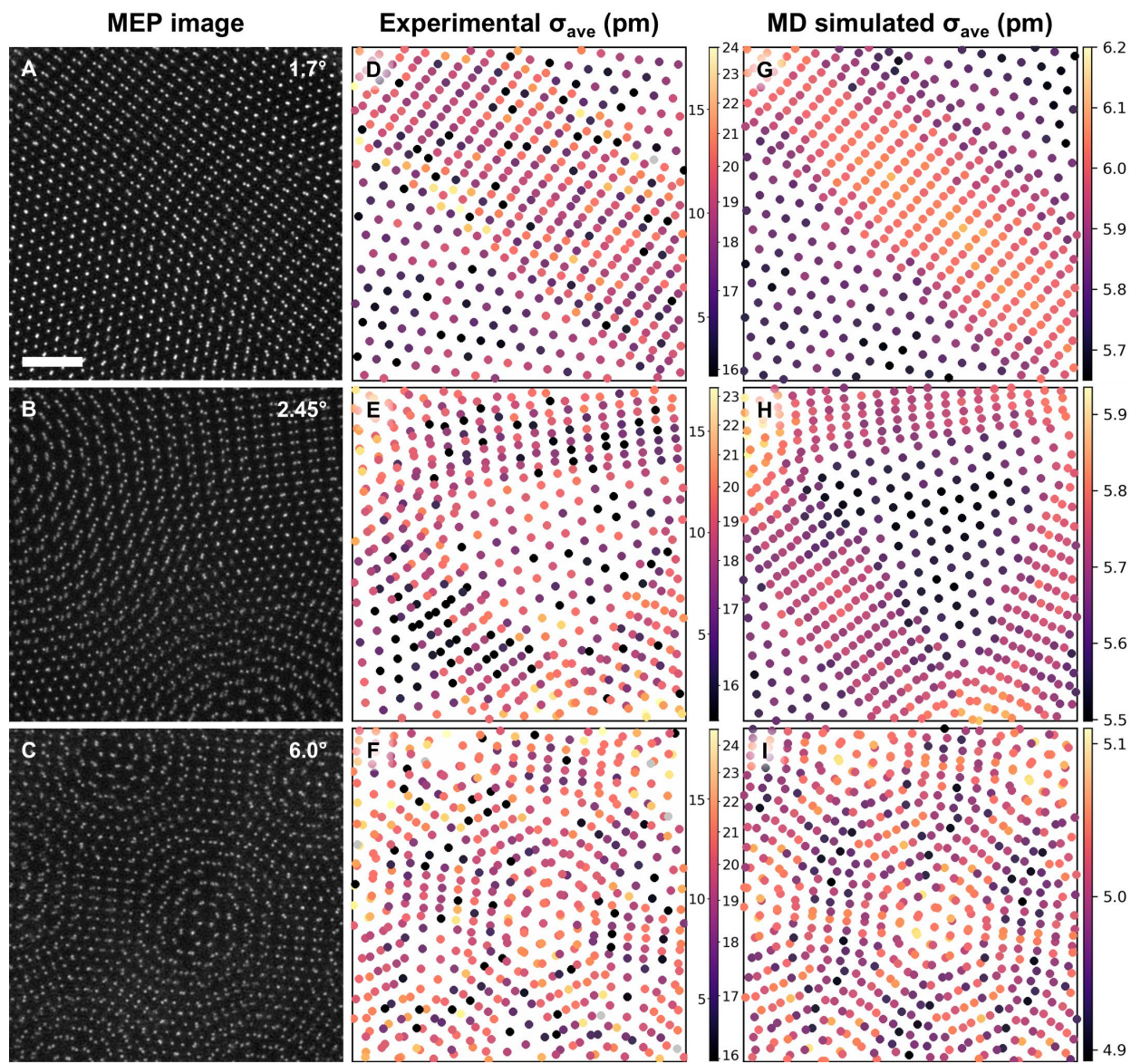

**Fig. 4 | Evolution of phasons with twist angle. A–C,** MEP image of 1.7°, 2.45°, and 6.0°-twisted bilayer WSe$_2$. **D–F,** Experimental, deconvolved average thermal vibration amplitudes σ$_{ave}$ (color bar left axis) and average Gaussian RMS atom width (color bar right axis) of single W atoms in **A–C,** respectively. **G–I,** MD simulated average RMSD of single W atoms of 1.7°, 2.45°, and 6.0°-twisted bilayer WSe$_2$, respectively. Scale bar = 1 nm for all panels.

## Materials and Methods

<u>Fabrication of twisted bilayer WSe$_2$</u>

Monolayers of WSe$_2$ are exfoliated using a modified gold assisted exfoliation method (*61*, *62*). A layered 100 nm gold and 500 nm copper film is deposited onto a flat silicon wafer. The film is lifted with double-sided Kapton® tape with only one side of adhesive exposed and a square window cut in the middle of the tape, where the tape acts as a handling layer. This gold-copper film with Kapton handling layer is then brought into contact with bulk WSe$_2$, purchased from 2D semiconductors®, with the gold side towards the crystal. The WSe$_2$ crystal, with metal film and the Kapton tape, is heated to 160°C, to ensure good adhesion. The metal film with Kapton tape handling layer is then carefully lifted off the bulk crystal with a monolayer of WSe$_2$ adhered to it. The film is placed on silicon/silicon dioxide wafer with 285 Å oxide thickness, where the metal film is carefully cut along the edges off the handling layer of Kapton tape. The wafer is then placed in a vacuum chamber to remove bubbles. The copper and the gold are subsequently etched using ammonium persulfate and potassium iodide solution, each etch followed by an active flow deionized water bath for 20 minutes.

<u>Transfer of twisted bilayer WSe$_2$ to TEM grids</u>

Twisted bilayer WSe$_2$ is constructed using a modified polycarbonate /Polydimethylsiloxane (PC/PDMS) lens pickup method (*63*). A large, clean flake of monolayer WSe$_2$ is identified. To ensure controllable twist angles, half of the flake is brought into contact with the PC/PDMS lens, then lifted off the wafer, ripping the flake in half. The stage is rotated to the desired twist angle, after which the lifted WSe$_2$ flake is brought into contact with the remaining flake on the wafer. Both layers are lifted off the wafer and placed onto a holey TEM grid, then heated to melt the PC sacrificial film, separating the heterostructure from the PC/PDMS lens. The TEM grid is then placed in multiple chloroform baths to remove PC. The TEM grid undergoes one additional isopropyl alcohol bath to remove carbonaceous residue and then air dried.

<u>Electron microscopy data acquisition</u>

The samples are imaged in a Thermo Fisher Scientific Themis Z aberration-corrected STEM operated at 80 kV. For both ADF-STEM imaging and 4D-STEM data acquisition, we use probe current 7–15 pA and convergence semi angle of 25 mrad. A 1$^{st}$ generation EMPAD detector is used to collect 4D-STEM data with 128 × 128 scan positions using a 0.43 Å scan step size and 1 millisecond dwell time. A defocused electron probe (-7.5 nm and 10 nm) is used during data acquisition. The camera length is 230 mm such that the maximum $k$-vectors collected is 2.5 Å$^{-1}$. A complete list of acquisition parameters is provided in Table S1.

<u>Ptychographic reconstruction</u>

We use a customized PtychoShelves package (*44*, *64*) to carry out the ptychographic reconstructions. The reconstruction uses the least-squares maximum-likelihood algorithm (*65*), mixed-states ptychography (*66*), probe variation correction (*67*), and scan position correction. The CBED patterns are padded to 256 × 256 pixels with zeros, corresponding to 9.79 pm real-space pixel size. Reconstructions converge within 500 to 800 iterations. The number of iterations needed depends on the data quality. A complete list of parameters used to reconstruct the experimental data can be found in Table S2.

Incorporation of extended depth of field (EDF)

After obtaining a $z$-stack of images from ptychographic reconstruction, we remove images in the stack containing mostly vacuum layers. The new $z$-stack is used to generate the EDF version of the ptychographic phase image using the *Extended Depth of Field* plug-in of FIJI (*55, 68*). We use the *complex wavelets* option for sharpness estimation with a filter length of 6 and number of scales 10; though *Sobel*, *variance*, and *real wavelets* options produce similar quality images.

Lattice dynamics and phonon calculations

Lattice dynamics calculations are performed using the empirical potential described by Naik *et al.* (*69*) to obtain the frequencies and eigenvectors of phonon modes. For the 1.7° twisted bilayer $WSe_2$ system, harmonic phonon calculations are conducted using the finite displacement method as implemented in PHONOPY (*70*). Initially, structural optimization of the hexagonal unit cell, consisting of 6846 atoms, is carried out in classical molecular dynamics (MD) simulation package LAMMPS (*71*) with energy and force tolerances set to $1E^{-11}$ and $1E^{-12}$ eV/Å, respectively. Subsequently, atoms within the optimized unit cell are displaced from their equilibrium position by 0.01 Å to generate the configurations required for evaluating the $2^{nd}$ order force constants. Based on crystal symmetry, PHONOPY (*70*) generated 41,076 distinct displaced configurations, for which the total energies and atomic forces were computed using LAMMPS (*71*).

Molecular-dynamics simulations

Classical MD simulations are performed via LAMMPS (*71*) using a hexagonal unit cell of 2.45-degree twisted $WSe_2$ system with 3282 atoms. The intralayer interactions are described by the Stillinger-Weber potential, while the interlayer interactions are modeled using the Kolmogorov-Crespi potential (*69*). The parameters for the aforementioned force fields for $WSe_2$ are taken from the Naik *et al.* (*69*) The structure is first equilibrated at 300 K for 100 ps using a Nose-Hoover thermostat with a timestep of 0.5 femtoseconds. To address the flying ice cube problem (*72*), a tether is applied to the center of mass of the system. Following the equilibration, the system is transitioned to microcanonical ensemble and relaxed for an additional 100 ps. Finally, the simulation is extended for 3 ns to generate atomic trajectories for subsequent analysis.

2D Anisotropic Gaussian fitting for measuring atom profiles in ptychography data

We first identify the atom positions coarsely by using a MATLAB package StatSTEM (*73*). We smooth the ptychographic phase images by applying Gaussian blur and thresholding before using a peak-finding algorithm to search for the local maxima in each image. This step identifies more than 95% of atomic columns in the twisted bilayers, and the misidentified atoms are corrected manually. Next, we use these rough positions as the initial guess for 2D anisotropic Gaussian fitting extract the size and shape of individual atoms.

Choosing the appropriate regions of interest (ROIs) before 2D anisotropic Gaussian fitting is critical because atoms are relatively close to each other, especially in the solitons and AA domains. Both rectangular and circular ROIs can lead to incorrect fitting due to tail of the nearby atom leaking into the ROIs. Therefore, we use a watershed image processing method to address this challenge. For each pixel in the image, we compare its pixel value with its 8 nearest neighbors (3 for corner pixels, 5 for edge pixels). If this pixel has the highest value, then it is marked as a "peak" (white points in Fig. S2B), and the nearest neighbors are included in a set that contains all pixels in the appropriate ROI (colored points in Fig. S2B) for Gaussian fitting. If its pixel value is lower than any of the nearest neighbors, then the peak search is moved to the highest value pixel in its nearest neighbors, and all of the pixels involved in this step are included in the ROI for Gaussian fitting. We repeat this procedure until all pixels are identified either as a peak or part of the ROI of a peak pixel. We then threshold to select the peaks corresponding to atoms rather than background noise. Only pixels as part of a ROI are used during 2D Gaussian fitting for each atom.

We use the following functional form of a 2D anisotropic Gaussian, shown in Eqn. S1:

$$G = A \cdot \exp(G_0) + B \qquad \text{Eqn. S1}$$

Here, $A$ is the amplitude factor, and $B$ is background. $G_0$ is shown in Eqn. S2:

$$G_0 = -\left( \frac{\left( (x - x_0) \cdot \cos(\theta) - (y - y_0) \cdot \sin(\theta) \right)^2}{2\sigma_x^2} + \frac{\left( (x - x_0) \cdot \sin(\theta) - (y - y_0) \cdot \cos(\theta) \right)^2}{2\sigma_y^2} \right) \qquad \text{Eqn. S2}$$

Here, $x_0$ and $y_0$ are the $x$-, $y$-coordinates of the center of the fitted Gaussian. $\sigma_x$ and $\sigma_y$ are the long and short-axis direction Gaussian root mean square (RMS) widths. $\theta$ is the angle between the long-axis and $x$-axis directions. Counterclockwise direction is defined as positive.

The ellipticity magnitude $\epsilon$ is calculated using Eqn. S3
$$\epsilon = \frac{\sigma_x}{\sigma_y} \qquad \text{Eqn. S3}$$

Ptychographic simulations

We first simulate 4D-STEM datasets of 1.7°- and 2.45°-twisted bilayer WSe$_2$ using the multislice method as implemented in Prismatic (74–76). The simulation parameters are chosen to match experimental acquisition conditions: 80 kV, 5 nm overfocus, 24.94 mrad convergence semi-angle, 0.429 Å probe scan step size, electron dose of $5.5 \times 10^5$ electron/Å$^2$, and 128 × 128 scan positions. We first bin the output CBED pattern by a factor of 2.6 and then crop it down to 128 × 128 pixels to match the experimental CBED patterns with a similar radius of the bright-field disk.

We simulate three scenarios: phonon-free, isotropic random phonons, and phonons from MD simulations. In the phonon-free case, no frozen phonons are added to the 4D-STEM multislice simulations. For isotropic random phonons, we input a thermal vibration amplitude, $\sigma_{\text{thermal}}$, of 5.48 pm for W atoms, 6.32 pm for Se atoms (57), and a total of 20 frozen phonons. For phonons from MD simulations, we sample atom coordinates at every 150 ps for a total of 3 ns (20 phonon configurations), simulate a 4D-STEM dataset for each phonon configuration, and average over CBED patterns for each phonon configuration.

We use the same PtychoShelves package for reconstructing the simulated 4D-STEM datasets. For ptychographic simulations, we use a single probe mode and no variable probe modes. The simulated CBED patterns are padded to $256 \times 256$ pixels with zeros. This corresponds to a 9.77 pm real-space pixel size, similar to the experimental value of 9.79 pm. Reconstructions converge within 500 to 800 iterations.

Deconvolution of thermal vibration amplitudes

After obtaining ptychographic phase images of the simulated 4D-STEM datasets, we extract the short-axis Gaussian RMS width of all single W atoms in the phonon-free, random-phonon, and MD-phonon cases using the same 2D anisotropic Gaussian fitting procedure as discussed previously. To perform deconvolution of thermal vibration amplitude, we first obtain an average short-axis Gaussian RMS width of single W atoms by fitting a 1D Gaussian function to the histogram of single W atom short-axis Gaussian RMS widths of phonon-free simulation (Fig. S7). This value, $\sigma_{\text{phonon-free}} = 15.5$ pm, is used as the basis to deconvolve the experimental thermal vibration amplitude $\sigma_{\text{thermal}}$ using Eqn. SEqn. S4 (44):

$$\sigma_{\text{thermal}} = \sqrt{\sigma_{\text{fit}}^2 - \sigma_{\text{phonon-free}}^2} \qquad \text{Eqn. S4}$$

We note that $\sigma_{\text{phonon-free}}$ varies with acquisition conditions and ptychographic reconstruction quality. Therefore, $\sigma_{\text{phonon-free}}$ is used to deconvolve experimental data in Figures 2 and 3. For data displayed in Figure 4, we scale $\sigma_{\text{phonon-free}}$ with respect to the information transfer in the Fourier transforms of the images. The Fourier transform of a ptychographic phase image is radially integrated to generate a spectrum. The peak position of the furthest identifiable peak is used to calculate $\sigma_{\text{phonon-free}}$ using Eqn. S5:

$$\sigma_{\text{phonon-free,new}} = \sigma_{\text{phonon-free}} \frac{D_{\text{FFT peak,new}}}{D_{\text{FFT peak}}} \qquad \text{Eqn. S5}$$

Here, $\sigma_{\text{phonon-free, new}}$ is the scaled value for a given experimental ptychographic phase image. $D_{\text{FFT peak}}$ is the position of the furthest identifiable peak in unit of reciprocal length in the simulated ptychographic phase image. $D_{\text{FFT peak, new}}$ is that of the given experimental ptychographic phase image.

Atom density plot fitting

The atom probability density function (Fig. 3B) is populated by the positions of the atom at each time step (0.5 fs) during the 3 ns of MD simulations. The 2D surface is sampled with 0.47 pm pixel size. When the atom position falls into a pixel, the count is increased by 1. A covariance matrix, $S$, is provided to describe the probability distribution of the atom position as shown below:

$$S = \begin{bmatrix} \sigma_x^2 & \rho\sigma_x\sigma_y \\ \rho\sigma_x\sigma_y & \sigma_y^2 \end{bmatrix}$$

Here, $\sigma_x$ and $\sigma_y$ are the Gaussian RMS widths corresponding to principal axes. $\rho$ is Pearson's correlation coefficient.

The long axis $\sqrt{\lambda_1}$ and short axis $\sqrt{\lambda_2}$, square root of eigenvalues $\lambda_1$ and $\lambda_2$, of the covariance matrix $S$ represent the thermal vibration amplitudes in orthogonal directions. $\lambda_1$ and $\lambda_2$, as well as the direction of ellipticity ($\theta$, angle between $x$-axis and long axis) are calculated from the covariance matrix using Eqn. S6–Eqn. S8:

$$\lambda_1 = \frac{\sigma_x^2 + \sigma_y^2}{2} + \sqrt{\left(\frac{\sigma_x^2 - \sigma_y^2}{2}\right)^2 + \left(\rho\sigma_x\sigma_y\right)^2} \qquad \text{Eqn. S6}$$

$$\lambda_2 = \frac{\sigma_x^2 + \sigma_y^2}{2} - \sqrt{\left(\frac{\sigma_x^2 - \sigma_y^2}{2}\right)^2 + \left(\rho\sigma_x\sigma_y\right)^2} \qquad \text{Eqn. S7}$$

$$\theta = \begin{cases} 0, & \rho\sigma_x\sigma_y = 0 \text{ and } \sigma_x \geq \sigma_y \\ \dfrac{\pi}{2}, & \rho\sigma_x\sigma_y = 0 \text{ and } \sigma_x \geq \sigma_y \\ \text{atan}\left(\dfrac{\lambda_1 - \sigma_x^2}{\rho\sigma_x\sigma_y}\right), & \text{else} \end{cases} \qquad \text{Eqn. S8}$$

The short-axis thermal vibration amplitude $\sqrt{\lambda_2}$ of each single W atom is plotted in Fig. 3E and Fig. 4G–I.

**Supplementary Text**

<u>Phonon dispersions in twisted bilayer WSe₂</u>

The phonon dispersion for the 1.7°-twisted bilayer WSe$_2$ system is presented along the high-symmetry path from Γ (0,0,0) to M (0,0,0.5) in Fig. S6. The three commonly seen acoustic modes (upper three modes, labeled LA, TA mode 1, TA mode 2) initially rise steeply from Γ with large group velocity but then become nearly dispersionless as they approach the zone boundary. More surprising are the two phason modes (labeled ultrasoft shear modes 1 and 2). These modes describe the sliding motion of the domain walls that separate commensurate regions (*i.e.*, AB regions) of the twisted bilayer and the lack of resistance to shear. Formally, the phason modes arise as a result of the coexistence of different types of stacking (*4*, *21*, *35*). For large twist angles, one layer is rigidly twisted with respect to the other (no internal relaxations), and the modes of the bilayer system are well described as the modes of isolated layers, largely decoupled from each other. In contrast, for small twist angles, internal relaxations result in the formation of the moiré superlattice with well-defined regions of alternate stacking separated by domain walls. The regions of inhomogeneous coupling lead to the mixing of otherwise low-frequency modes of the isolated layers, inducing the formation of the shear modes. In the limit of an incommensurate twist, the shear modes would be acoustic in nature. Their frequencies approach zero at Γ, reflecting the invariance of the moiré to horizontal shifts of one layer relative to the other. In this limit, the shear modes can be understood as the gapless Goldstone modes of the soft, incipient moiré superlattice (*4*). For intermediate twist angles, the shear modes are also present but exhibit imaginary (negative) frequencies (*5*, *35*) associated with the structural transition from the unrelaxed rigid regime observed at large twist angles, to the relaxed moiré regime observed at small twist angles. As a result of the commensurate twist angle, they show a gapped, optical nature with near zero dispersion at Γ.

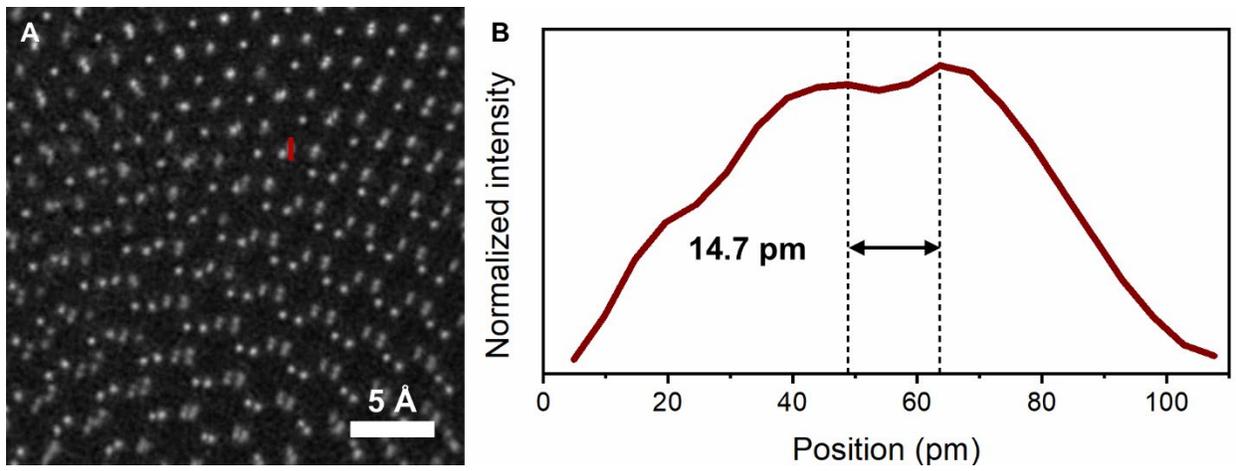

**Fig. S1 | Smallest projected atomic spacing resolvable with multislice electron ptychography (MEP). A**, MEP phase image of 2.45°-twisted bilayer $WSe_2$. **B**, Line profile showing a resolvable projected atomic spacing of 14.7 pm.

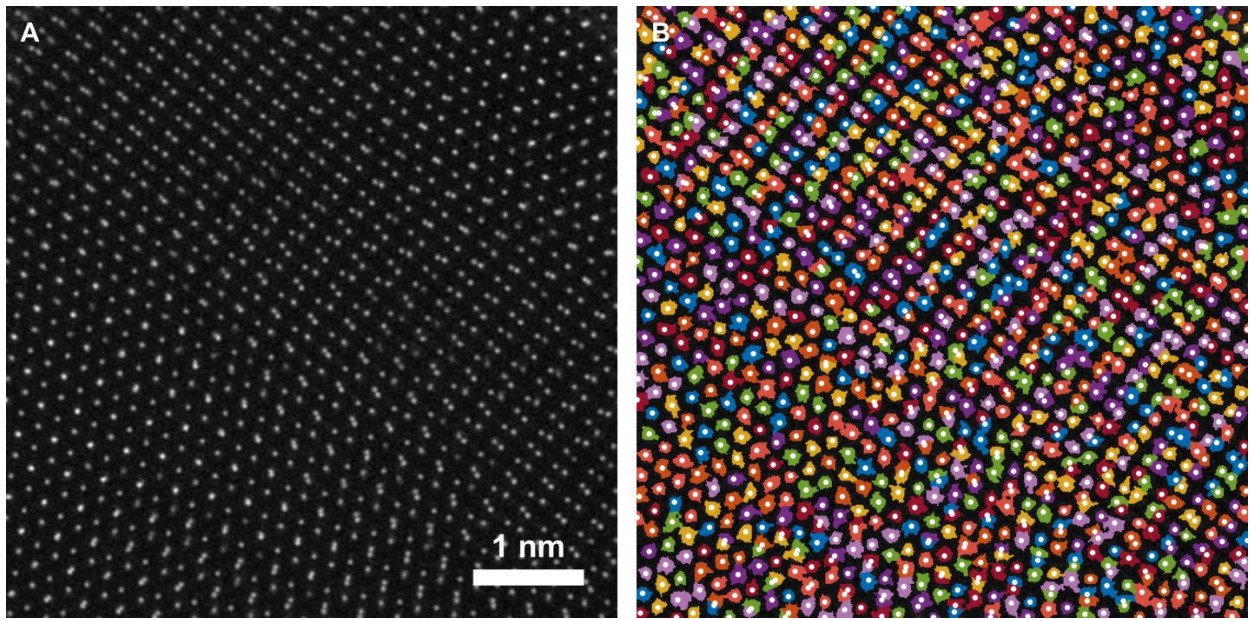

**Fig. S2 | Determining regions of interests (ROI) for Gaussian fitting using a watershed method. A**, MEP phase image of 1.7°-twisted bilayer WSe$_2$. **B**, ROIs used for 2D Gaussian fitting plotted on the same image shown in **A**. Each ROI is labeled by one color (the same color may be used to label a nearby atom due to finite number of colors used). White dots mark atom pixel positions as initial guess for 2D Gaussian fitting.

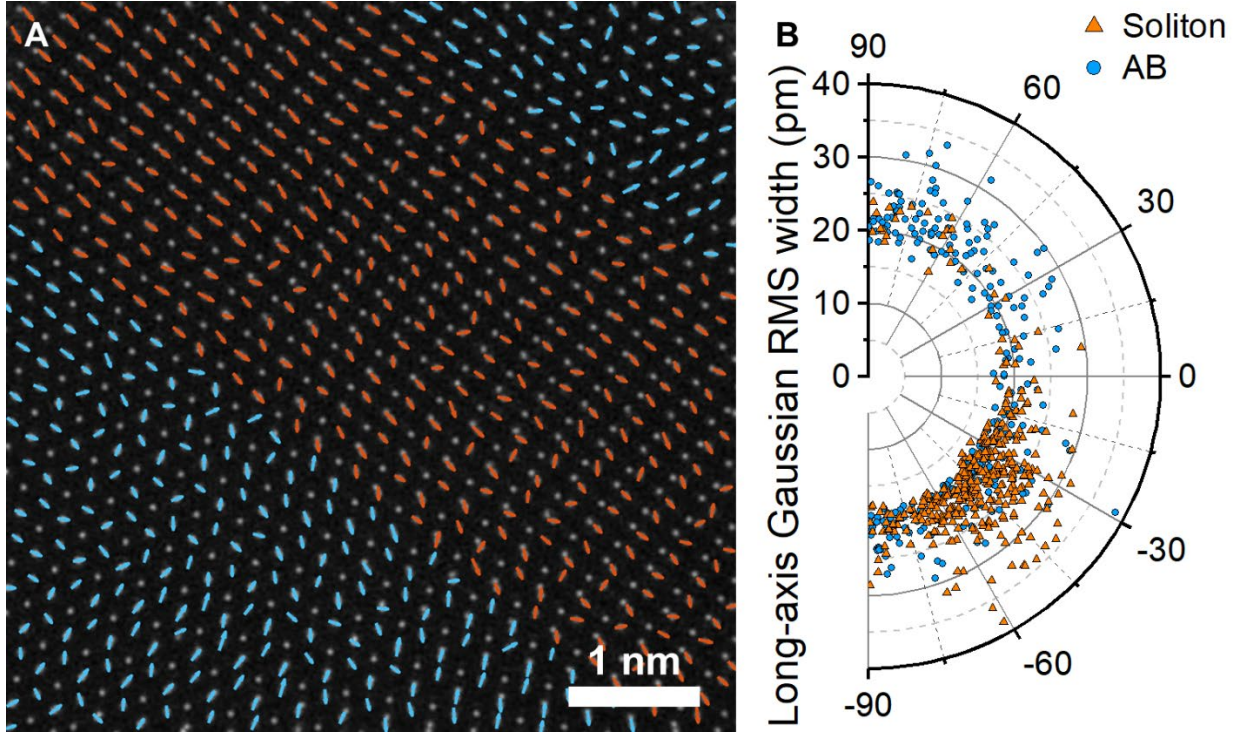

**Fig. S3 | Ellipticity analysis of Se atomic columns. A**, Vector field map of anisotropic phonon vibration overlaid on MEP phase image of 1.7°-twisted bilayer WSe$_2$. Orange and blue arrows indicate the relative magnitude and direction of the ellipticity of Se atomic columns in soliton and AB regions, respectively. **B**, Polar plot of long-axis Gaussian RMS width of Se atomic columns. Se atomic columns behave similarly to single W atoms shown in Fig. 2E. We display results of the single W atoms in the main text to avoid potential artifacts arising from Se pair splitting due to local tilt or shear.

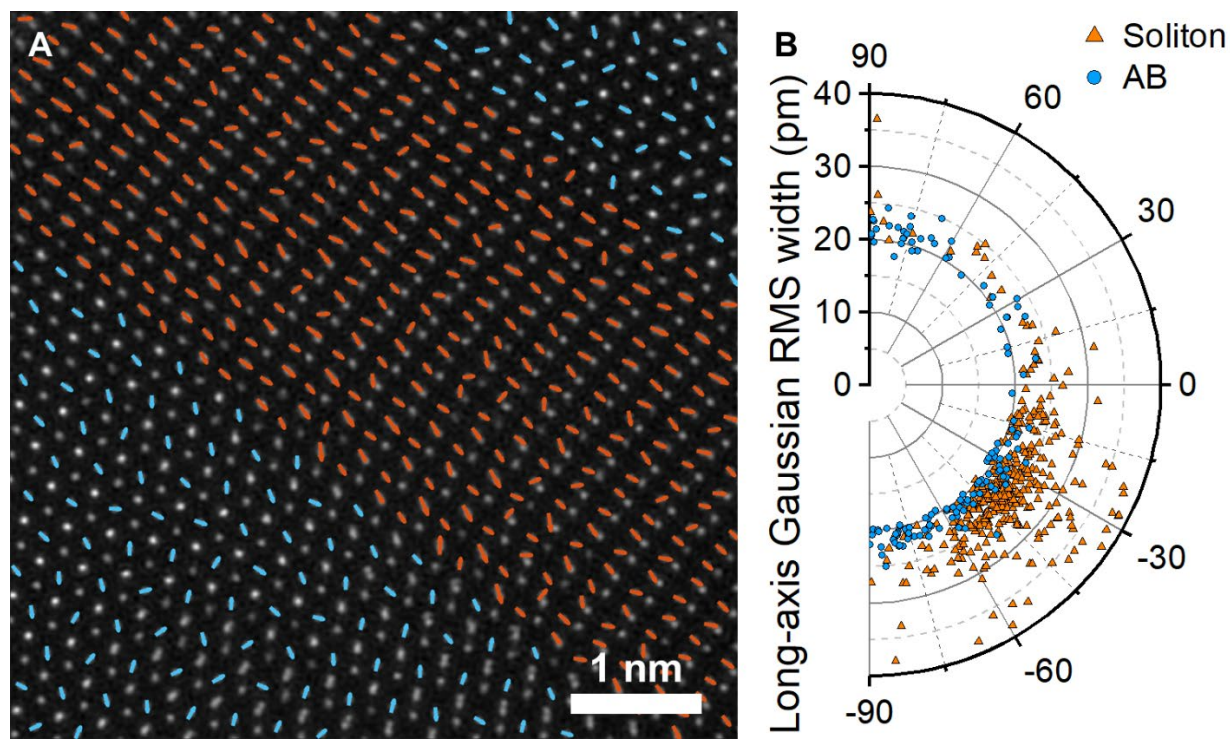

**Fig. S4 | Ellipticity analysis on conventional MEP projection (without EDF). A**, Vector field map of anisotropic phonon vibration overlaid on MEP phase image of 1.7°-twisted bilayer WSe$_2$ without EDF. Orange and blue arrows indicate the relative magnitude and direction of the ellipticity of all single W atoms in soliton and AB regions, respectively. **B**, Polar plot of long-axis Gaussian RMS width of single W atoms.

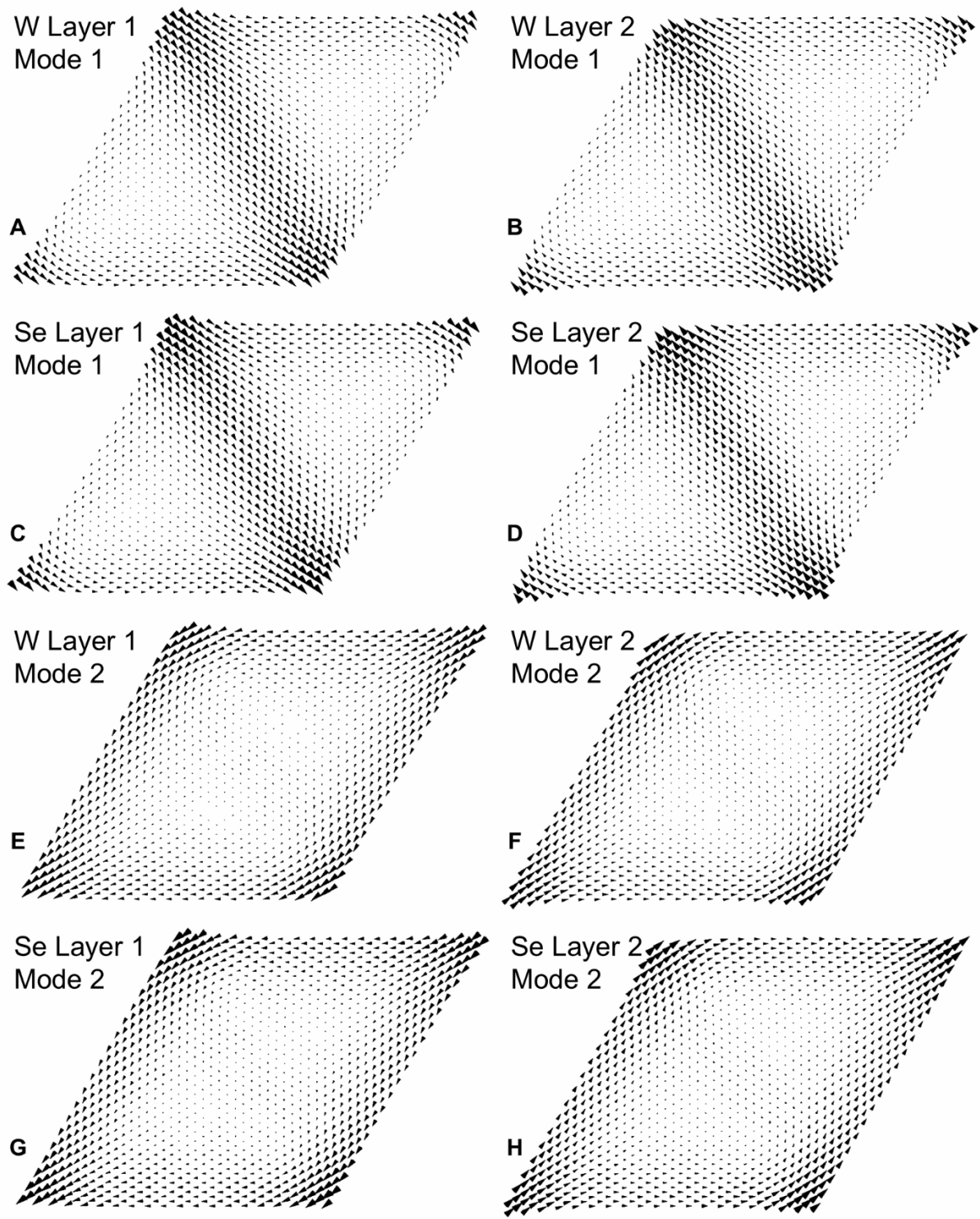

**Fig. S5 | Eigenmodes of two lowest energy moiré phasons in 1.7°-twisted bilayer WSe₂. A–D,** Eigenvector maps of top and bottom layer W atoms and Se atoms for Eigenmode 1 (lowest energy mode). **E–H,** Eigenvector maps of top and bottom layer W atoms and Se atoms for Eigenmode 2 (second lowest energy mode).

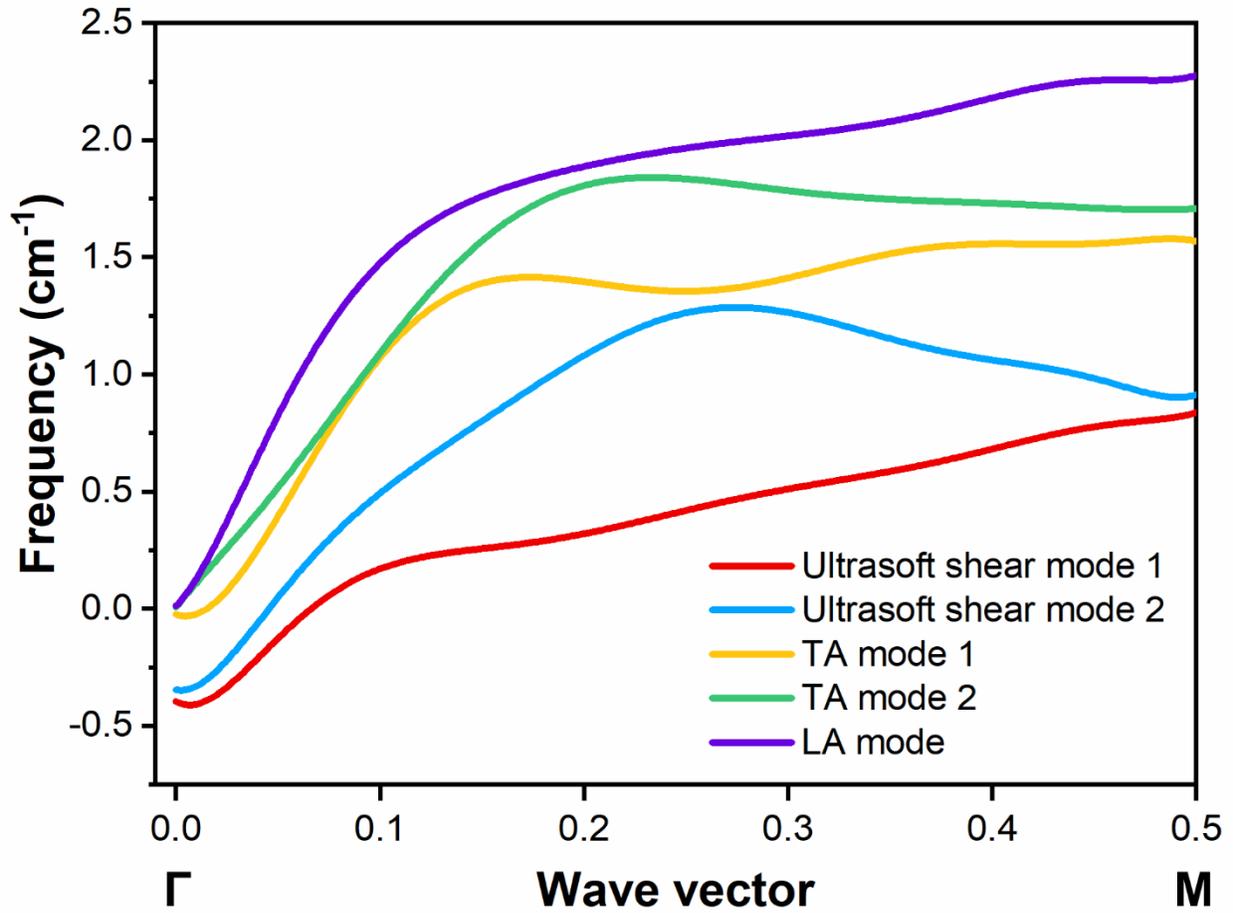

**Fig. S6 | Phonon dispersion in 1.7°-twisted bilayer WSe₂.** To emphasize the ultrasoft shear modes (phason modes), the phonon dispersion is shown for the first five modes along the high-symmetry path from Γ (0,0,0) to M (0,0,0.5). The plot identifies five modes: three acoustic modes, including the longitudinal acoustic (LA) mode and two transverse acoustic (TA) modes, along with two phason modes (ultrasoft shear modes 1 and 2).

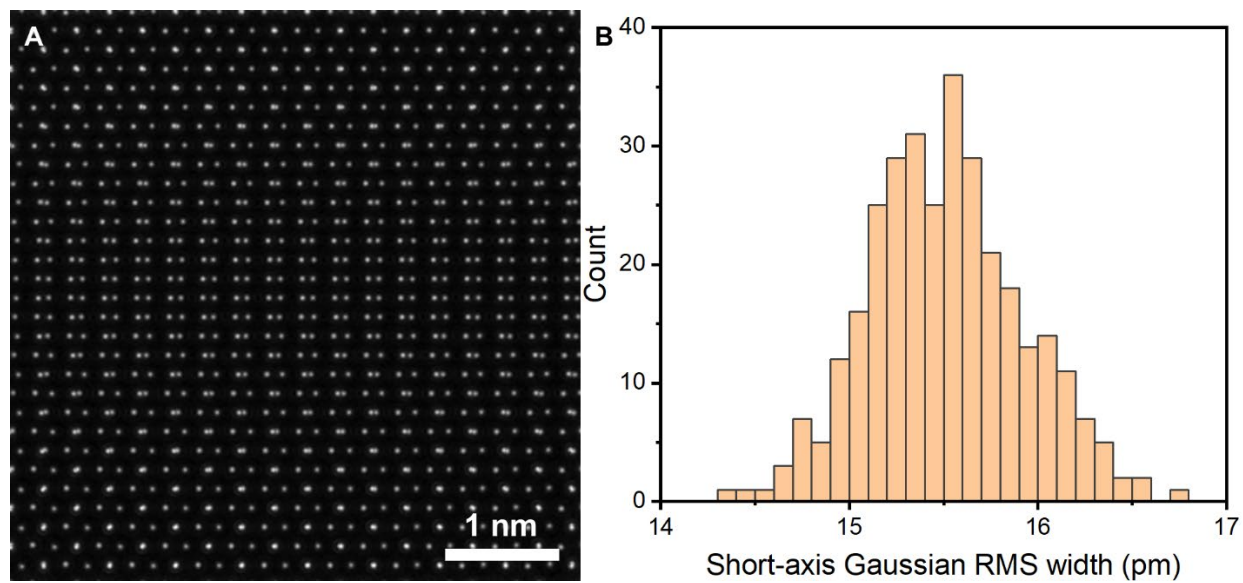

**Fig. S7 | Phonon-free ptychographic simulation for deconvolution. A**, Simulated MEP phase image of 1.7°-twisted bilayer WSe$_2$. **B**, Histogram of short-axis Gaussian RMS width of all single W atoms. 1D Gaussian fit returns center of distribution at 15.5 pm. We use this mean value to deconvolve the contributions of ptychography and the atomic potential from thermal vibrations.

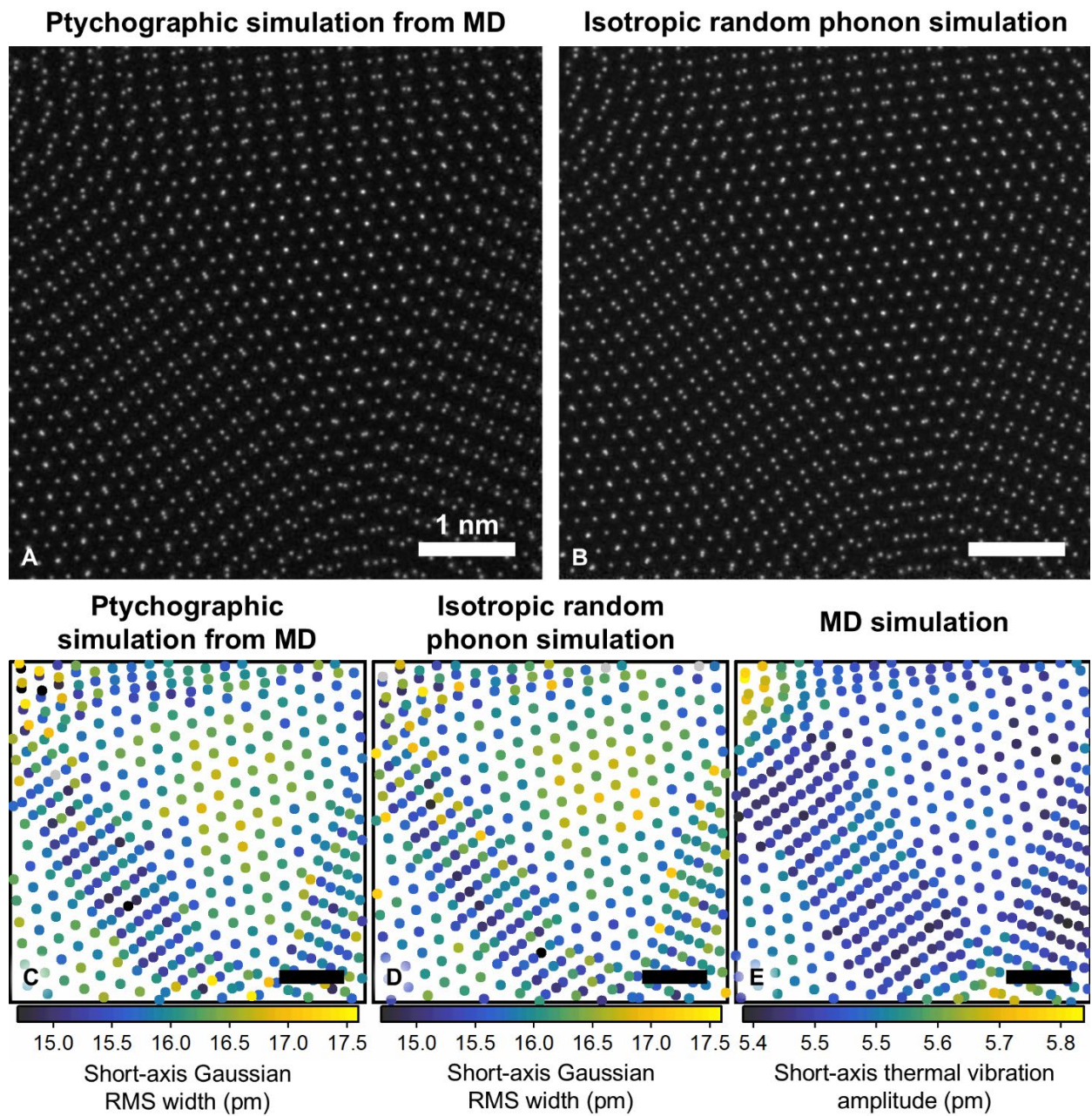

**Fig. S8 | Short-axis maps from ptychographic simulations. A,B,** Simulated MEP phase images of 2.45°-twisted WSe$_2$ bilayers of the same region using phonon configurations from **A,** MD simulations and **B,** isotropic random phonons. **C,D,** Short-axis Gaussian RMS width maps of single W atoms in **A** and **B**, respectively. **E,** MD simulated short-axis thermal vibration amplitudes of single W atoms. Scale bar = 1 nm for all panels. Comparison between **C,D** shows that ptychography simulations using phonon configurations from MD exhibit increased short-axis Gaussian RMS width near AA sites, while ptychography simulations with isotropic random phonons do not.

**Table S1. List of data acquisition parameters**

| | Fig. 2A | Fig. 3A | Fig. 4C |
|---|---|---|---|
| *Convergence angle* | 25 mrad | | |
| *Max. k-vector* | 2.5 Å$^{-1}$ | | |
| *Scan step size* | 0.43 Å | | |
| *# scan positions* | 128 × 128 | | |
| *CBED pattern size* | 128 | | |
| *Defocus* | 7.5 nm underfocus | 10 nm overfocus | 10 nm overfocus |
| *Electron dose* | $2.4 \times 10^5$ e$^-$/Å$^2$ | $6.8 \times 10^5$ e$^-$/Å$^2$ | $5.1 \times 10^5$ e$^-$/Å$^2$ |

**Table S2. List of ptychographic reconstruction parameters**

| | Fig. 2A | Fig. 3A | Fig. 4C |
|---|---|---|---|
| *Presolve CBED size** | | 64, 128 | |
| *CBED zero padding* | | 256 | |
| *# probe states* | | 10 | |
| *# variable probe modes* | | 2 | |
| *Regularization* | | 0.1 | |
| *Total thickness* | | 20 Å | |
| *# slices* | 8 | 12 | 8 |
| *Scan position correction on #iteration* | | 1 | |
| *Total #iterations* | | 800 | |

*Presolve means running ptychography reconstruction using a reduced CBED patterns first to improve convergence. For example, when using CBED size 64, only the central 64 × 64 pixels in the CBED patterns are used for ptychographic reconstructions. The probe and object of this reconstruction are used as the initial guess for the next round of reconstruction, where all 128 × 128 pixels of the CBED patterns are used for the reconstruction. The probe of this reconstruction is used as the initial guess for reconstructions where zeros are padded to the CBED patterns.

**Movie S1. Movie of lowest energy phason eigenmode (Eigenmode 1)**. Top view of atomic motions of W (gray) and Se (blue) atoms within one moiré supercell of 1.7°-twisted bilayer WSe$_2$ corresponding to eigenmode vector map in Fig. 2F, Fig. S4, and Fig. S5.

**Movie S2. Movie of second lowest energy eigenmode (Eigenmode 2)**. Top view of atomic motions of W (gray) and Se (blue) atoms within one moiré supercell of 1.7°-twisted bilayer WSe$_2$ for Eigenmode 2 in Fig. S5.